\documentclass{sig-alternate-2013}

\permission{Copyright is held by the International World Wide Web Conference Committee (IW3C2). IW3C2 reserves the right to provide a hyperlink to the author's site if the Material is used in electronic media.}
\conferenceinfo{WWW 2015,}{May 18--22, 2015, Florence, Italy.} 
\copyrightetc{ACM \the\acmcopyr}
\crdata{978-1-4503-3469-3/15/05. \\
http://dx.doi.org/10.1145/2736277.2741684}

\usepackage[
  pass,
]{geometry}

\usepackage{url}
\usepackage{color}

\usepackage{amsmath}
\usepackage{ifthen}
\usepackage{graphics}
\usepackage{graphicx}
\def\etal{\textit{\emph{et al}.}}

\newcommand{\reminder}[1]{}

\usepackage{cite}
\usepackage{graphicx}
\usepackage{times}

\def\sneq{{\tiny \mbox{$\neq$}}}
\def\subsetneq{\ \lower.5ex\hbox{$\stackrel{\subset}{\small \sneq}$}\ }

\newcommand{\tworows}[2]{
\begin{tabular}{@{}l@{}}#1\\#2\end{tabular}
}

\newcommand{\omt}[1]{}

\newcommand{\graf}[3]{
\ifthenelse
 {\equal{#1}{h}}
  {\includegraphics[height=#2\textheight]{#3}}
  {\includegraphics[width=#2\textwidth]{#3}}
}

\newcommand{\xhdr}[1]{\subsection*{\bf #1}}

\usepackage{array}
\newcolumntype{C}[1]{>{\centering}m{#1}}

\def\appsrand{{APPS$_{RAND}$}}
\def\appspop{{APPS$_{POP}$}}

\begin{document}

\title{The Lifecycles of Apps in a Social Ecosystem}

\numberofauthors{4}

\author{
Isabel Kloumann\\
       \affaddr{Cornell University}\\
       \affaddr{Ithaca, NY}\\
       \email{imk36@cornell.edu}
\and
Lada Adamic\\
       \affaddr{Facebook Inc.}\\
       \affaddr{Menlo Park, CA}\\
       \email{ladamic@fb.com}
\and
Jon Kleinberg\\
       \affaddr{Cornell University}\\
       \affaddr{Ithaca, NY}\\
       \email{kleinber@cs.cornell.edu}
\and
Shaomei Wu \\
       \affaddr{Facebook Inc.}\\
       \affaddr{Menlo Park, CA}\\
       \email{shaomei@fb.com}
}
\date{\today}

\maketitle
\begin{abstract}
Apps are emerging as an important form of on-line content, and they combine aspects of Web usage in interesting ways --- they exhibit a rich temporal structure of user adoption and long-term engagement, and they exist in a broader social ecosystem that helps drive these patterns of adoption and engagement.  It has been difficult, however, to study apps in their natural setting since this requires a simultaneous analysis of a large set of popular apps and the underlying social network they inhabit.

In this work we address this challenge through an analysis of the collection of apps on Facebook Login, developing a novel framework for analyzing both temporal and social properties.  At the temporal level, we develop a retention model that represents a user's tendency to return to an app using a very small parameter set.  At the social level, we organize the space of apps along two fundamental axes --- {\em popularity} and {\em sociality} --- and we show how a user's probability of adopting an app depends both on properties of the local network structure and on the match between the user's attributes, his or her friends' attributes, and the dominant attributes within the app's user population.  We also develop models that show the importance of different feature sets with strong performance in predicting app success.
\end{abstract}

\vspace{0.05in}
\noindent
{\bf Categories and Subject Descriptors:}
H.2.8 [{\bf Database Management}]: Database applications---{\em Data mining}

\vspace{0.0in}
\\
\mbox{~}
\vspace{0.0in}

\section{Introduction}
There is, or is likely soon to be, a webservice or app for virtually
every component of modern life.  They are diverse and ubiquitous;
they constitute both a backdrop and chronicle of everyday experience.
And they represent a broad change in overall patterns of Internet use ---
both the research community and the media have increasingly
begun discussing the ``appification of the Web'' 
\footnote{\url{https://sites.google.com/site/appweb2012/}}.
Yet empirical opportunities to consider them as a complete
ecosystem have been limited, and as a result we still know very little
about the population structure of apps ---
their inherent diversity, their lifecycles, and the ways
in which users engage with them.

The high-level characteristics of app engagement as a form of Web use
are still the subject of much discussion and refinement, but
certain properties emerge independent of any one particular
app's functionality --- these include {\em temporal} properties,
based on long-running patterns
of individual usage and engagement over time,
and {\em social properties},
in which an individual will typically
be a user of many apps with overlapping functionality,
in a broader social environment that is
bootstrapped to create within-app social activity.

To address these issues,
we study the collection of apps on Facebook Login,
making use of anonymized aggregate daily usage logs of the
apps and web services accessible through this mechanism.
We undertake our analysis on two levels of scale --- the individual level,
focusing on the properties of user behavior over time and
in relation to other users;
and the app level, modeling the overall usage level of the app
and the social structure on its users.

At the temporal level, we develop a user retention model, showing how
with a small number of parameters we can approximate the probability
that a user who adopts an app at time $t$ will continue to be using
it at a future time $t + \Delta$.
The model exposes the ways in which usage decay has a time-dependent
component, and provides us with a compact set of parameters representing
a particular app's engagement profile that can then be used
in higher-level tasks.
When we consider the app's user population as a whole, we are led to natural
lifecycle modeling and prediction questions --- given an app's
history up to a given point in time, how well can we predict its number
of users going forward? Interesting recent work of Ribeiro~\cite{Ribeiro:2014}
considered this question using time-series data for several
large Web sites; we show how a broad range of feature categories ---
including our derived retention parameters, together with
individual characteristics of
the app's users and the social network structure on its full user
population ---
can lead to strong prediction results
across a wide diversity of apps.

At the level of app social structure, we show how the space
of all the popular apps on Facebook Login can be organized in
a two-dimensional representation whose axes correspond to
{\em popularity} --- the raw number of users --- and {\em sociality} ---
the extent to which users of the app have friends who are also
users of the app.
This representation exposes certain global organizing principles
in the full app population, including a pair of complementary
``frontiers'' to the space --- one containing apps whose sociality
is relatively fixed independent of their popularity, and one in
which the sociality of the app's user population
is not much greater than that of a random set of Facebook users
of comparable size.

Finally, we perform an analysis of social characteristics at the
individual user level, analyzing the Facebook users who are one step away
from an app in the social network --- a set we can think of as the
``periphery'' of the app, containing people who are not yet users of
the app, but have friends who are users.
For a person in an app's periphery, we can attempt to predict
future adoption of the app based on individual characteristics and
network structure.
We find that apps are diverse in the way in which the structure
on a user's friends is related to adoption probabilities,
and we find an interesting effect in the interactions among
individual characteristics: a user's probability of adopting an app
depends on the three-way relationship among their own attributes,
the attributes of their friends who use the app, and the
modal attributes of the full population of app users.

\section{Data}
The data for this study comes from anonymized logs of Facebook Login daily
activity, collected between January 2009 and June 2014.  Facebook Login is a
secure way for Facebook users to sign into their apps without having to create
separate logins.

The various analyses in this paper required different slices of these logs,
both considering the observation window and the apps being observed. Table 1
summarizes the different subsampled data sets that will be referred to
throughout this work. The data for this study has granularity of one day; that
is, we have logs about whether an individual uses a specific app on each day.
All user level data is de-identified.

\begin{table}[htpb!]
\begin{tabular}{|p{0.6in}|p{0.8in}|p{0.6in}|p{0.76in}|}
\hline
\hline
tag & \tworows{selection}{criteria} & time period & size \\
\hline
\appsrand & \tworows{random $\propto$}{ DAU(2014-06)} & \tworows{Jan. 2014 -}{Jun. 2014} & 83,000 apps\\
\hline
\appspop & \tworows{most popular by}{MAU(2013-06)} & \tworows{Jan. 2009 -}{Jun. 2014} & \tworows{2,319 apps}{$1.4\times 10^9$ users}  \\
\hline
\hline
\end{tabular}
\caption{Summary of data sets considered in this paper. DAU and MAU refer to
Facebook Daily Active Users and Monthly Active Users, respectively. Our initial
overview analyses consider APPS$_{RAND}$, while our subsequent and in-depth
analyses consider APPS$_{POP}$ and, as occasion permits, various subsets of it
(APPS$_{POP\{X\}}$). Unless otherwise noted, subsampling in this work is done
on apps, not on users.}

\label{tab:data}
\end{table}

The frequency with which the Facebook Login service is called, and hence daily
activity is registered, depends on several factors. Web-based activity relies
on authentication tokens that expire on the order of hours, while mobile apps
can optionally request tokens that are valid for days, provided the user does
not change their password. For some apps, we do see a periodic activity,
typically 7 days apart, consistent with longer-term authentication tokens being
refreshed. This periodicity is a small effect relative to the overall activity,
as we show below.  This is likely because other activity, such as posting
updates or retrieving public profiles or friend lists, again requires
reconnecting.  
Therefore, Facebook Login provides a reasonable proxy of daily use of the app.
It allows us to characterize the app's adoption and retention.

\section{Social Properties of Apps}

\subsection{Popularity and sociality}
One question that has been raised previously is how big of a role the social
network plays in the adoption of apps. This parameter has been inferred
indirectly by Onnela and Reed-Tsochas~\cite{onnela2010spontaneous} in their
study of the very early adoption of Facebook apps.  It is also estimated in the
model proposed by Ribeiro ~\cite{ribeiro2014modeling}, where individuals
can drive their friends' adoption and re-engagement.

However, these prior studies did not directly measure whether app adoption was
in fact correlated on the network, and so we turn to this task presently.  In
particular, we would like to place apps in a low-dimensional space that can
provide a view for how they are distributed across the social network of users.

To do this, we begin with two basic definitions
\begin{itemize}
\item We say that the {\em popularity} of an app, denoted $p(x)$,
is the probability
that an individual selected uniformly at random from
Facebook's population is a user of the app.
\item We say that the {\em sociality} of the app, denoted $p(x|y)$,
is the probability that a member of Facebook is a user of the app given that
they have at least one friend using the app.
\end{itemize}

Studying the distributions of $p(x)$ and $p(x|y)$, and how they
are jointly distributed across apps, allows us to ask a number of
questions.  In particular, how socially clustered is the app?
And how does it depend on the type of app, or characteristics of
the app's users?

Note that if $p(x|y)$ is very high for an app, it means that its user
population in a sense ``conforms'' to the structure of the
underlying social network.

Moreover, $p(x|y)$ can in principle be high even when
$p(x)$ is low --- this would correspond to an app that is popular
in a focused set of friendship circles, but not on Facebook more broadly.
On the other hand, if $p(x|y)$ is not much more than $p(x)$,
then it says that users of the app are spread out through the
social network almost as though each member of Facebook independently flipped
a coin of bias $p(x)$ in order to decide whether to become a user
of the app --- there would be no effect of the social network at all.


\begin{figure}[hT!]
\includegraphics[width=1.1\linewidth]{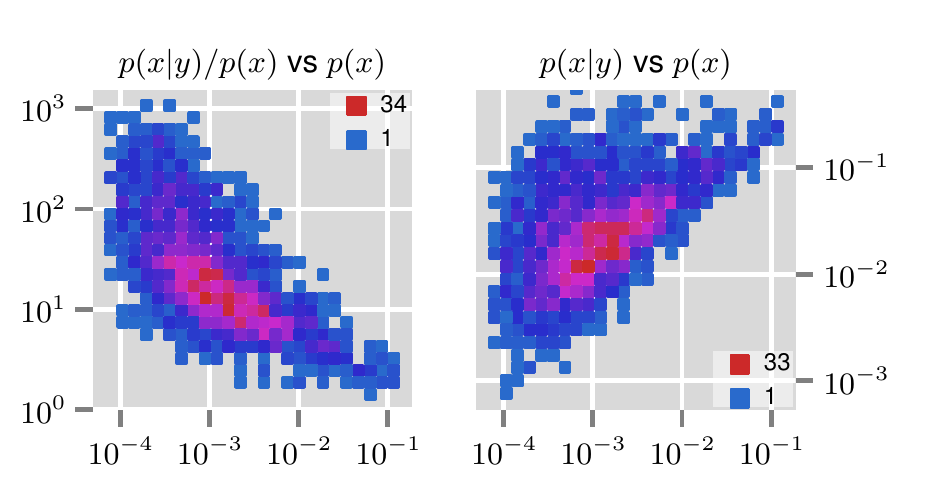}
\includegraphics[width=0.5\linewidth]{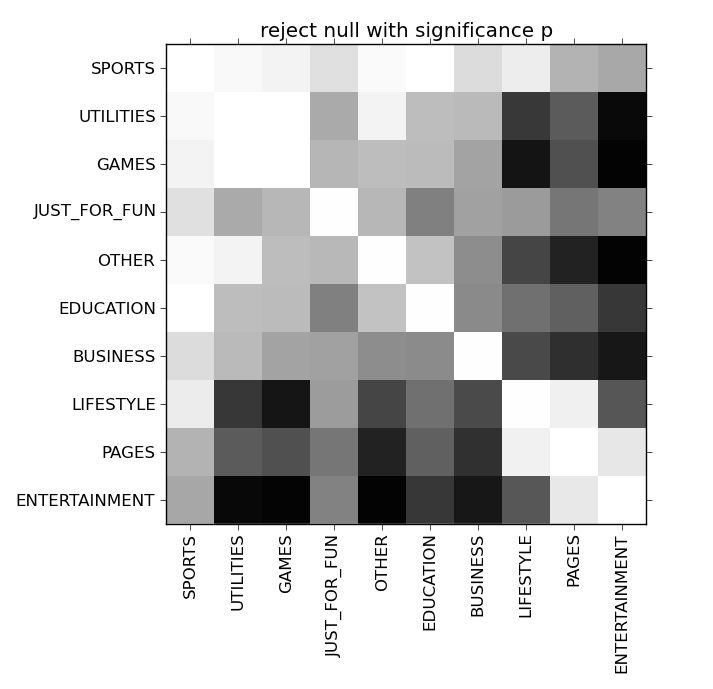}
\includegraphics[width=0.5\linewidth]{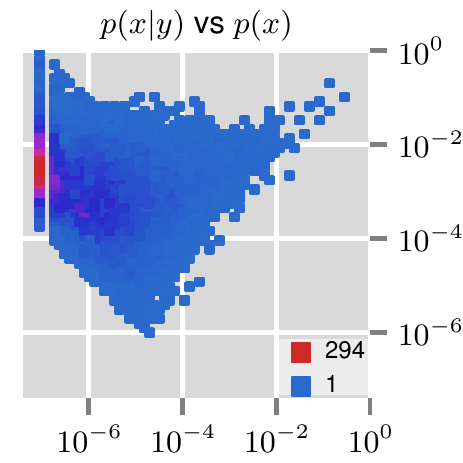}
\caption{App sociality. \textbf{ Top left:} Horizontal axis is app popularity,
and vertical axis is the relative increase in adoption likelihood for people
who have friends who also use the app. \textbf{ Right panels:} Horizontal axis
is app popularity, vertical axis is app sociality. The colors represent the
number of apps falling within the given bin. The lower right panel uses
\appsrand, while the other three panels use \appspop\ (see Table \ref{tab:data}
for details). The labeled colors indicate the relative frequencies of
observations in each bin, such that the lowest values have been normalized to
1. \textbf{ Bottom left:} Matrix indicating the $p$-values for the two sided
Kolmogorov-Smirnov test comparing the distributions $p(x|y)/p(x)$ for apps within each pair of the nine listed categories. White indicates a lower $p$-value and
black indicates a higher one.}
\label{fig:sociality_heatmaps} \end{figure}

\xhdr{Plotting apps in popularity-sociality space}

An appealing feature of this pair of parameters is that it provides
a natural two-dimensional view of the space of all popular apps
on Facebook.
We show this view in Figure \ref{fig:sociality_heatmaps} ---
a heat map showing the density of apps at each possible
(discretized) pair of values $(p(x), p(x|y))$.

We see in Figure \ref{fig:sociality_heatmaps} that the apps fill out a
wedge-shaped region in the $p(x)$-$p(x|y)$ plane, and it is informative to
understand what the boundaries of the region correspond to.  First, note that
if the social network had no relationship to app usage, we would see the
diagonally sloping line $p(x) = p(x|y)$; in the plot this corresponds to a line
that lies slightly below the diagonal lower boundary of the points in the heat
map.  Thus, there exists a frontier in the space of apps that is almost
completely asocial --- those apps that lie parallel to this diagonal line ---
but essentially no apps actually reach the line; even the most asocial apps
exhibit some social clustering.
We see this in the approximately horizontal top boundary of
the points in the heat map --- this is a frontier in the space of
apps where knowing that a person $x$ has a friend using the app
gives you a fixed probability that $x$ uses it, independent of
the app's overall popularity on Facebook.
The location of this horizontal line is interesting, since it
provides an essential popularity-independent value for the maximal
extent of social clustering that we see on Facebook.

Note that the wedge-shaped region in a sense has to come to
a point on the right-hand side, as $p(x)$ becomes very large:
once an app is extremely popular, there is no way to avoid
having pairs of friends using it almost by sheer force of numbers.
And given the crowding of app users into the network,
there is also no way for the extent of social clustering to
become significantly larger than one would see by chance.

The third, lower left boundary of the wedge is a manifestation of the Facebook
friend limit of 5000: if a user is friends with someone who uses the app, at
least 1/5000 of their friends use it. We approach this limit in the far left
hand of this Figure with apps that have two users, one with 0 other friends
using the app and the other with 1 of their nearly 5000, combining to approach
the lower limit of $1/(2 * 5000)$. The lower bound decreases as $1/(n * 5000)$
as the number of users, $n$, increases.

\begin{figure}[tb!]
\includegraphics[width=\linewidth]{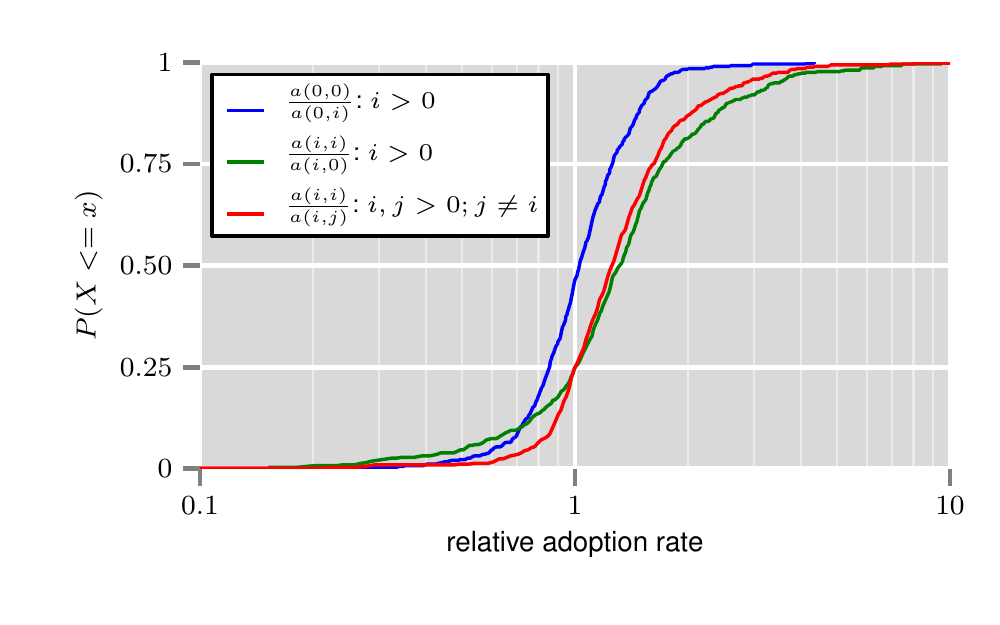}
\caption{
Relationship between national identity of potential app adopters, that of their
current user friends, and the likelihood of their adopting the app. 
Blue: Relative adoption rates when a potential user and their current user friend are from the majority to when potential user
is in majority and current user friend is in minority.  
Green: Relative rates when potential user and current user friend are in same minority to when potential user
is in minority and current user friend is in majority. 
Red: Relative rates for when potential user and current user friend are in same minority to when they
are in different minorities. 
The blue curve indicates that when a potential user is from the majority country, their current user friend could be from either the majority or a minority and they are still
equally likely to adopt an app more often as less. In contrast, when the
potential user is from a minority country, in 75\% of apps they adopt more frequently
when their current user friend is from the same country as them.
} \label{fig:countryadopt}
\end{figure}

\subsection{Analysis of Social Neighborhoods}

We saw in the previous section that app adoption can be localized in the social
network. But what is the mechanism by which the app is adopted by friends?  Is
homophily driving both friendships and adoption of specific apps based on
interests?  Or is the primary mechanism one of exposure, where having even just
a single friend who has installed the app now gives an opportunity to become
familiar with the app and subsequently adopt it?

To answer these questions, we observe the Facebook friendship graph at time
$t$. For each app, we consider everyone on Facebook who has friends using the
app, but who has not used the app themselves by $t$. Are there features of the
individual and their friends that will predict whether
the individual will adopt
the app at some point in the future?

\xhdr{One-node neighborhoods}

We begin with a question about homophily and its relation to app usage.
Consider a Facebook user $A$ who does not currently use the app, and
suppose that has exactly one friend $B$ who uses the app (that is, $A$ has
between 1 and 5000 friends on Facebook, but for our purposes here, exactly one
of those friends is an app user). We choose some attribute on users (for
example nationality, or age); we let $f(A)$ and $f(B)$ denote the value of this
attribute for $A$ and $B$, and we let $f^*$ denote the modal (or median) value
of the attribute across all app users.

Now the following question arises.  Suppose that $A$ is different
from the typical user in this attribute, in the sense that
$f(A) \neq f^*$.  Is $A$ more likely to adopt the app if the friend $B$
is similar to $A$, or if $B$ is similar to the typical app user?
This is a basic question about the role of individual similarity
in adoption decisions --- if we're studying potential users who are
outside the target demographic, is it more effective for their
app-using friends to be similar to them, or similar to the target
demographic?

We study this for nationality as an attribute in
Figure \ref{fig:countryadopt} --- given an app, we index nationalities so that
the most common nationality among users of the app is labeled $0$,
and other nationalities are labeled by values $i > 0$.
We now say that $a(i,j)$ is the adoption probability of a user $A$
who has one friend $B$ using the app, with $f(A) = i$ and $f(B) = j$.
(Note that $f^* = 0$ according to our notation, since $0$ is the most
prevalent nationality in the app.)
The figure shows that for a considerable majority of apps, we have
$a(i,i)/a(i,0) > 1$, indicating a clear aggregate tendency for the
question in the previous paragraph: a user $A$ is more likely to adopt the
app in general when $A$'s one friend using the app is similar to $A$,
not to the typical app user.
In contrast, when $f(A) = f^*$, the ratio $a(0,0)/a(0,i)$ is balanced
around $1$, so there is no clear tendency in adoption
probabilities between the case $f(B) = f^*$ and $f(B) \neq f^*$:
for users who have the modal attribute value, the attribute value of
their friend does not have a comparably strong effect.

\begin{figure}[tb!]
\includegraphics[width=\linewidth]{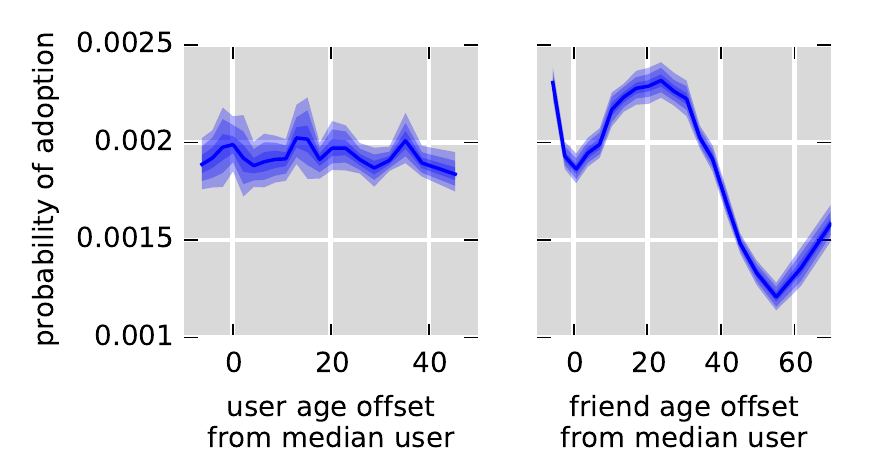}
\caption{The probability that a user adopts the app given that they have one
friend using the app. as a function of (left) the friend's age offset from the
median and (right) the user's age offset from the median.  The left plot
indicates no apparent relationship between the age of the friend and that the
user adopts. 
In contrast, the right plot illustrates that young users and users
who are aged between 10 and 30 years above the median age are more likely to
adopt. Users who are more than 40 years older than the median age are less
likely to adopt. The probabilities were binned by age into 20 equally populated
bins and the reported adoption probabilities are bootstrap estimates.  The
thick central line is the median bootstrap estimate of the mean, while the
three outer bands indicate the 68\%, 95\%, and 99.7\% confidence-intervals.  }
\label{fig:ageadopt} \end{figure}
This style of question gives us a way of analyzing individual
attributes in general, and we find that attributes differ in
the way this effect manifests itself.
For example, when we consider age as an attribute (in place
of nationality), we see (Figure \ref{fig:ageadopt})
that if $A$ has a friend $B$ using the app,
the age of $B$ has very little correlation with $A$'s probability
of adopting.
However, the age of $A$ is related to the adoption probability ---
users $A$ who are much older or younger from the median age are relatively
more likely to adopt the app if they have a friend who uses the app, compared
to individuals who are near the median age themselves.

\xhdr{Two- and three-node neighborhoods}
\label{sec:twothreenodes}

We can also use the population of apps, and the adoption
decisions that people make about them, to address a
recurring recent question in the literature on on-line diffusion.
Given an
individual $A$ who is not currently using an app, but who
has $k$ friends $B_1, B_2, \ldots, B_k$ using it,
how does $A$'s adoption probability depend on the
pattern of connections among these $k$ friends?
Is $A$ more likely to adopt if there are many links among these friends,
or very few?

Past work has suggested that the answer can depend on the
adoption decision being studied.
Consider the results observed by Backstrom
\etal~\cite{Backstrom2006group} with LiveJournal data, where conversion
probability increases with the connectedness
of one's friends, and contrast them with those observed by Ugander \etal
~\cite{Ugander:2011} with Facebook e-mail invitations, where, for a fixed
neighborhood size, one's probability of conversion strictly increased with the
number of independent components.  In both cases the result is non-obvious, as
there is no a priori mathematical reason that the effect should be monotone
with connectedness of one's neighbors.

Given the diverse answers arising in prior work, and the consequent
suggestion that the result depends on the adoption decision,
we consider how adoption probabilities vary with neighbor
connectivity across a large sample of the most popular apps.

\begin{figure}[h!]
\includegraphics[width=\linewidth]{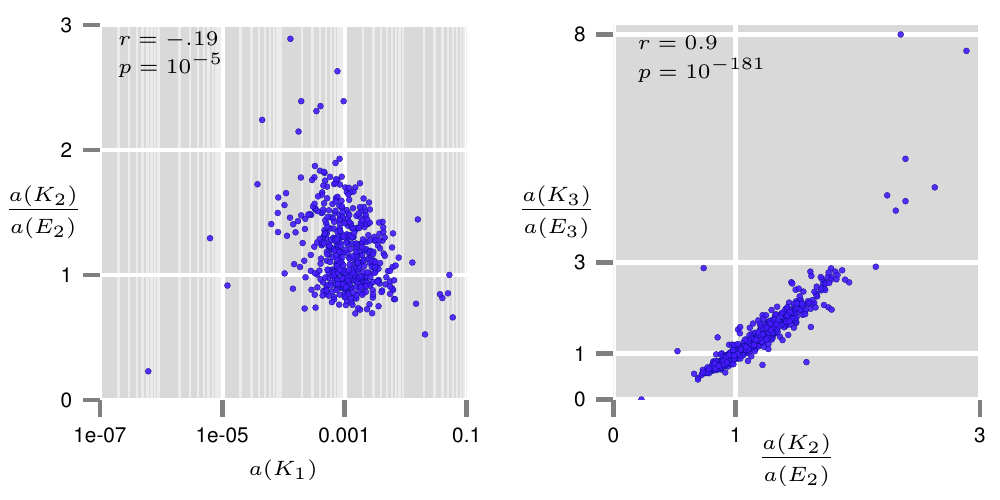}
\caption{Right: the baseline rate of adoption given that a user has two friends
using (horizontal) and the ratio of probability of adoption given friends are
connected to probability given that friends are not connected (vertical).  Apps
above the line $y=1$ exhibit the same trend as LiveJournal adoptions, and those
below follow the trend observed in Facebook adoptions.  \textbf{Left:}
Closed to open conversion ratio for two-node neighborhoods (horizontal) and
three node neighborhoods (vertical). Apps in the upper right quadrant follow
the LiveJournal trend for two- and three-node neighborhoods, while apps in the
lower left follow the Facebook trend. The correlation between these rates is
0.98 with $p<<0.01$, and there is a stark deficiency of apps in the diagonal
quadrants.}
\label{fig:k2k3}
\end{figure}

We begin by considering the question just for two-node user 
neighborhoods,$\ $ asking$\ $ it$\ $ separately$\ $ for$\ $ six$\ $hundred$\ $ apps$\ $ from \\\appspop: given that
a non-user $A$ has exactly two friends using the app at time $t$, how does
$A$'s adoption probability depend on the presence or absence of an
edge between $A$'s two friends? Note again that these users may have any number of Facebook friends between 2 and 5000, but that only two of those friends can be app users.
We explore this question in Figure \ref{fig:k2k3}.
We find that both possibilities --- higher or lower adoption probability
with the presence of an edge --- occur in roughly comparable proportions
across the population of apps, suggesting that at the level of two nodes,
both possibilities are indeed prevalent.

\begin{figure}[hbt!]
\includegraphics[width=0.5\linewidth]{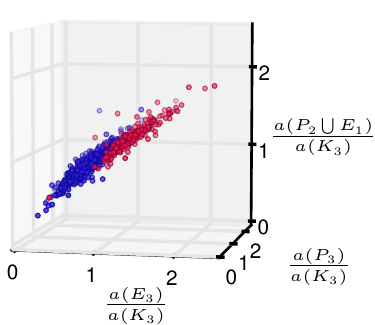}
\includegraphics[width=0.5\linewidth]{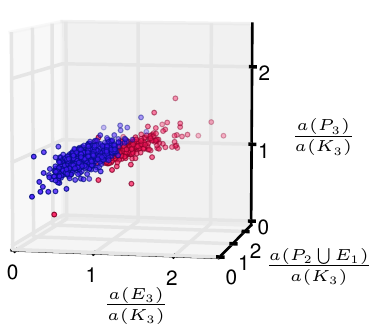}
\caption{
Two views of the same 3D point cloud: apps positioned according to the ratios
$a(E_3)/a(K_3)$, $a(P_2 \bigcup K_1)/a(K_3)$, and $a(P_3)/a(K_3)$, and colored
such that blue apps have $a(K_2) / a(E_2) \ge 1$, and red have $a(K_2) / a(E_2)
< 1$.  All adoption rates are reported relative to the rates for when a
friend's user friends are a clique. Red apps have higher adoption rates with
lower connectivity in the two-node graphs, and we see a near perfect
correspondence in this trend for each possible combination of connectivity in
three node graphs; this is demonstrated by the blue and red points falling
naturally on either side of 1 in all three dimensions.  In the left hand view
the vertical extent of the cloud demonstrates the natural variation in relative
adoption rates when a friend's user friends form two components (y-axis)
compared to three (x-axis). In contrast, in the right hand view of this point
cloud we observe more limited variation in relative adoption rates when the
user friends are connected in one component. These differences are reflected in
the Pearson correlation coefficients of 0.94 between $a(E_3)/a(K_3)$ and $a(P_2
\bigcup K_1)/a(K_3)$, and 0.57 between $a(E_3)/a(K_3)$ and $a(P_3)/a(K_3)$ ($p
< 10^{-10}$ in both cases). }
\label{fig:a3d}
\end{figure}

\begin{figure}[hT!]
\includegraphics[width=0.5\linewidth]{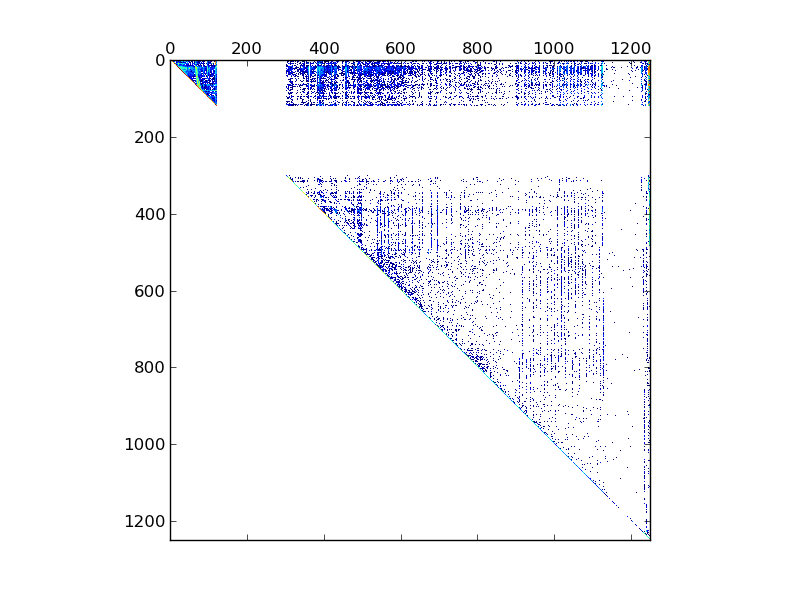}
\includegraphics[width=0.5\linewidth]{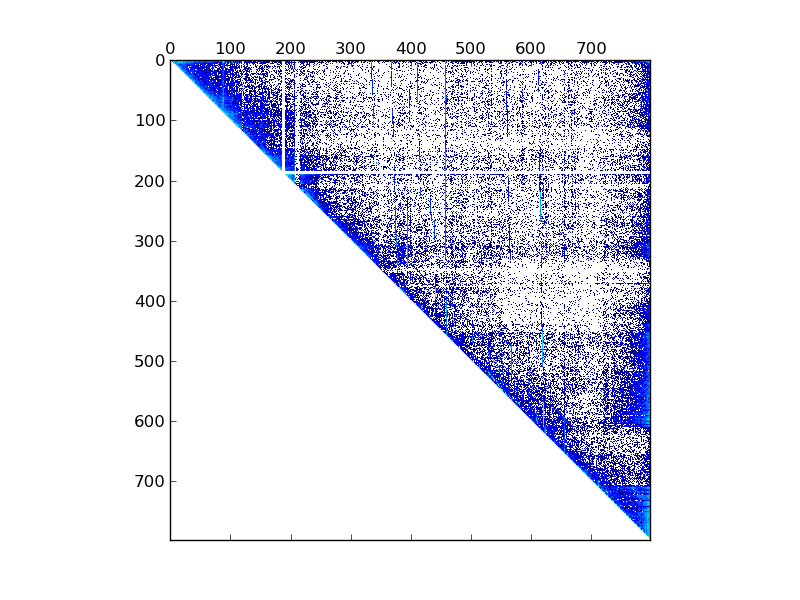}
\includegraphics[width=0.5\linewidth]{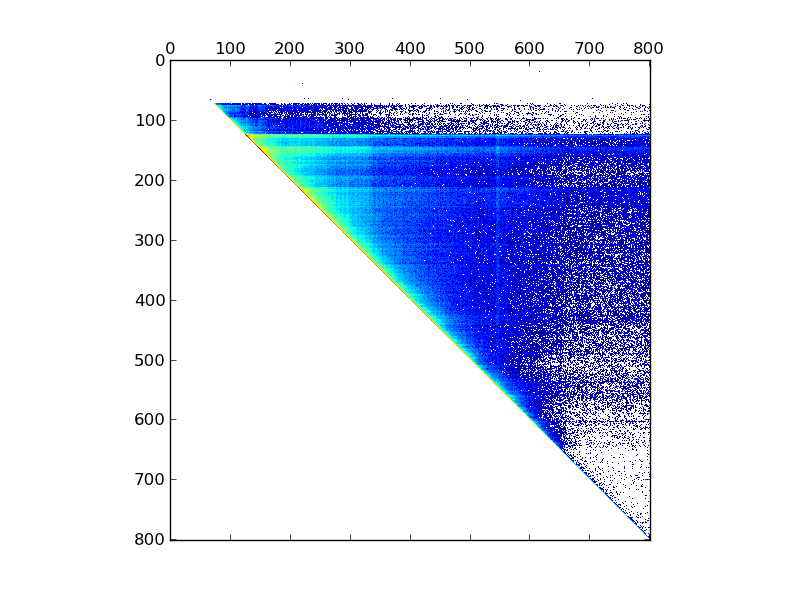}
\includegraphics[width=0.5\linewidth]{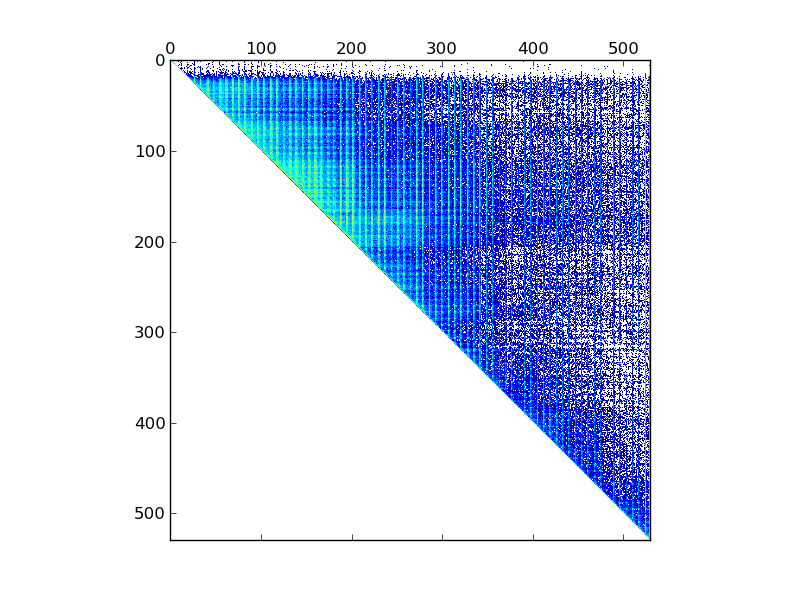}
\includegraphics[width=0.5\linewidth]{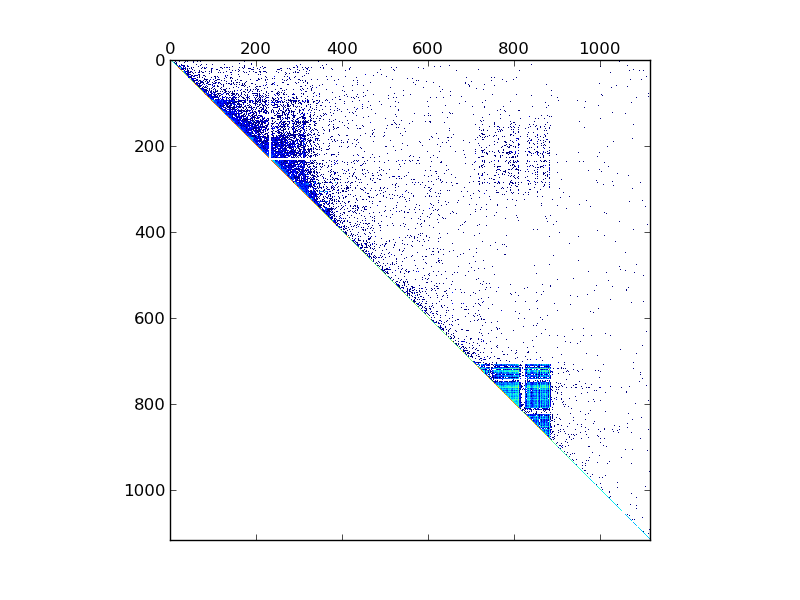}
\includegraphics[width=0.5\linewidth]{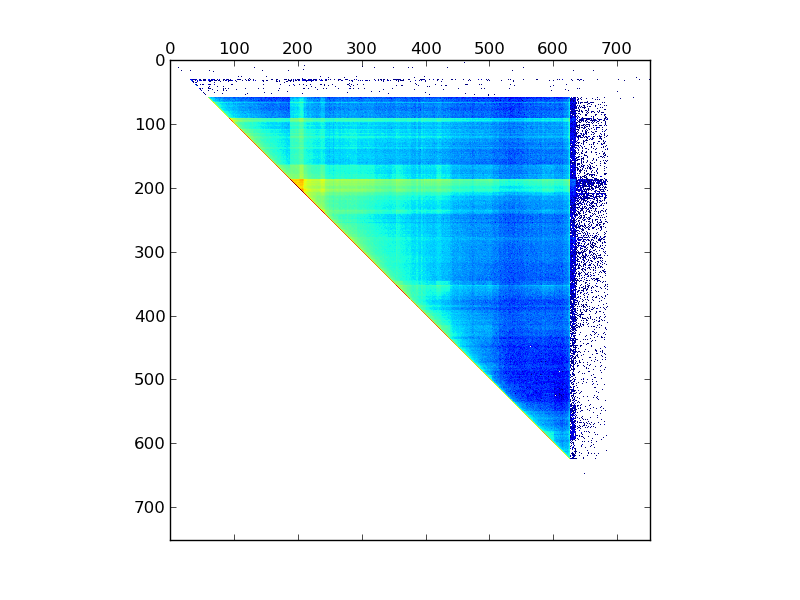}
\includegraphics[width=0.5\linewidth]{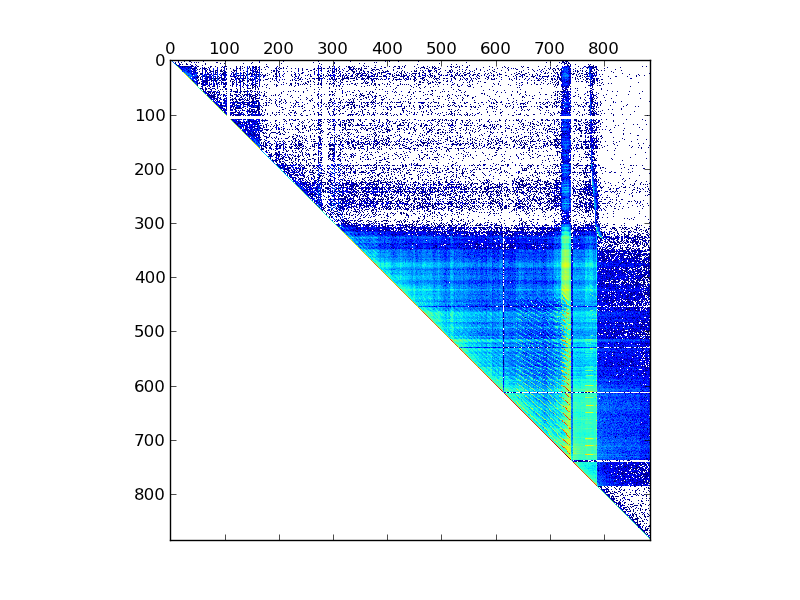}
\includegraphics[width=0.5\linewidth]{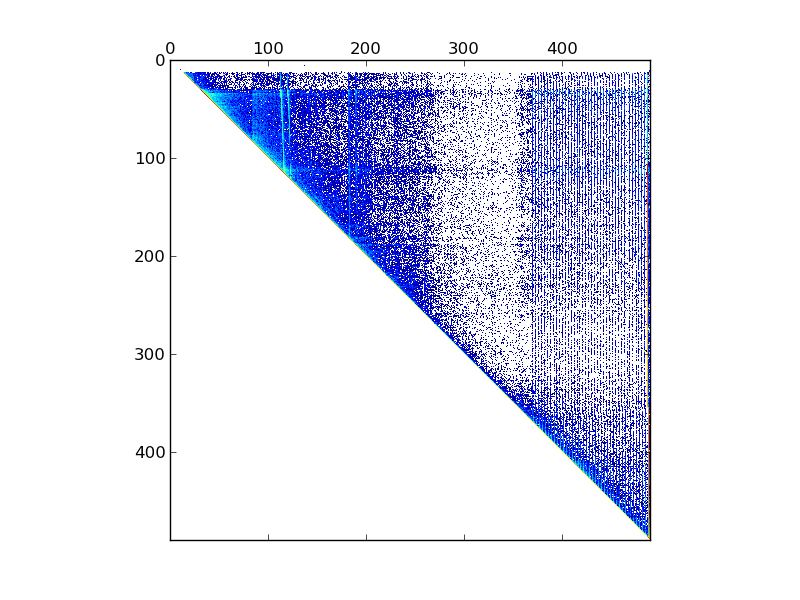}
\includegraphics[width=0.5\linewidth]{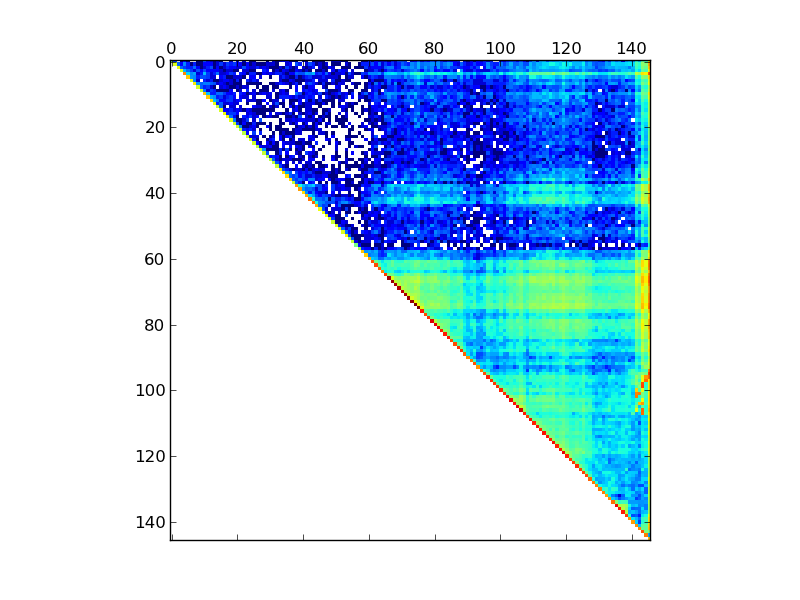}
\includegraphics[width=0.5\linewidth]{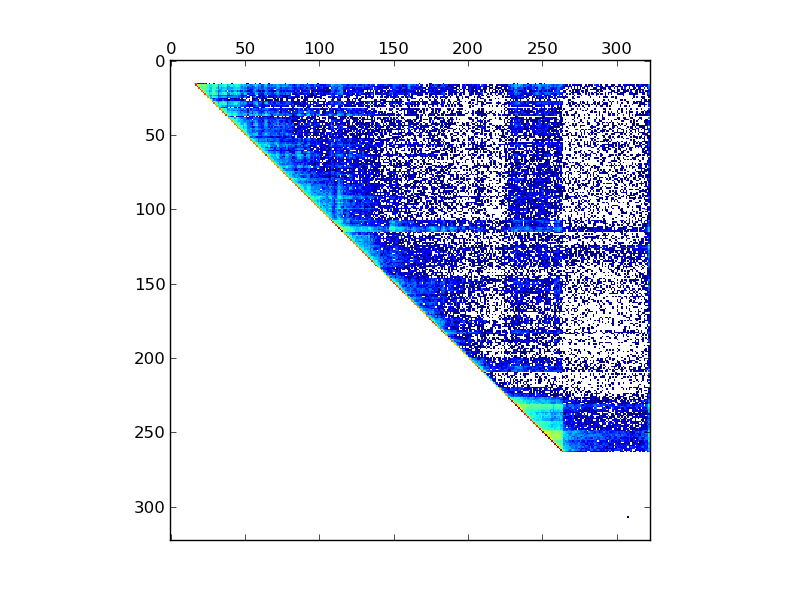}
\caption{App adoption and departure dynamics. Heat maps of aggregate first
and last login times for users of several examplar apps. The y-axis corresponds
to the date of the first login and the x-axis to the last. The concentration is
from blue (few) to yellow (many). Bright
yellow and green horizontal or vertical bands correspond to periods of rapid
adoption and departure, respectively. The color scale increases in density from white to blue, then yellow, then red.}
\label{fig:mau_heatmaps}
\end{figure}

As often happens in the analysis of phenomena on small social subgraphs,
we begin to see some rich structure emerge in the question when
we move to three-node neighborhoods.
For a graph $G$, we let $a(G)$ denote the adoption probability of user $A$
when $A$'s neighbors induce the graph $G$, and note that
on three nodes there are four possible graphs:
the complete graph $K_3$,
the three-node path $P_3$,
the single-edge graph $P_2 \cup E_1$, and
the empty graph $E_3$.

We find (Figure \ref{fig:k2k3})
that the ratio $a(K_3)/a(E_3)$ covaries closely with
$a(K_2)/a(E_2)$ across the set of apps --- in other words,
when the adoption probability of an app is higher for a connected
pair of friends, it is also higher for a connected triplet of friends.
This indicates how properties of two-node neighborhoods
provide strong information about the properties of larger neighborhoods ---
and it is an empirical regularity of the adoption decisions rather
than a strictly mathematical one, since the property for three nodes
does not follow from the property on two nodes.
We find similar regularities in the fact (Figure \ref{fig:a3d})
that when adoption probabilities are higher for $a(E_3)$ relative
to $a(K_3)$, they are also higher for the next-sparsest graph
$a(P_2 \cup E_1)$.
And we see comparatively much less variation between $a(P_3)$ and
$a(K_3)$, which is consistent in interesting ways with the finding
of Ugander \etal \cite{Ugander:2011} that neighborhood topologies
inducing the same number of connected components tended to lead
to similar adoption probabilities.

\section{Temporal patterns in apps}
In addition to learning how the adoption of apps depends on friendship ties,
we'd like to characterize
the app's ability to retain those users who have adopted.
These temporal features of an app's evolution
hold some of the keys to its success.

To get a sense for what the retention of users looks like at a global
level, we show the evolution of usage for a sample of apps in
Figure \ref{fig:mau_heatmaps}.
The images in this figure are heat maps in which the cell in the
$(i,j)$ entry records the density of users who first used the app at time $i$
and last used it at time $j$.
These maps thus show periods of heavy recruitment
as horizontal bands --- days when many individuals first started using the app.
While there are
a few vertical bands, denoting a narrow period of time
when many users were last active,
there is a clear concentration along the diagonal of many users departing soon
after their first login, while some, located off-diagonal,
remain active much longer.
There are also sudden increases and decreases in density, as apps
became more or less popular.
With this type of global view of the diversity in usage and retention patterns,
we next turn to modeling these retention patterns;
our goal is to extract parameters from such a model
to use as features in predicting an app's future engagement.

\subsection{Retention model}

For an app to have long term success we expect that it needs to
maintain a relatively high level of user retention.
We would like to have a model of retention that characterizes not only
whether an individual will log into the app the very next day, but for any
day subsequent to their first login.

Past work on retention modeling (e.g. ~\cite{YangICWSM2010,Karnstedt2011,Debeauvais2011}) has been focused on a particular product/activity (mostly online games), trying to predict users' continous engagement with a wide selection of features, many of them are domain-spefic and computationally expensive. To study thousands of apps and billions of users, we want to propose a model that is easy to compute and highly generalizable.


\xhdr{Simplest model: exponential decay}

We start with the population of newly installed users, $n(0)$, and assume that
at every time step each user has a constant probability of leaving, $x_0$. This
mechanism gives rise to exponential decay:

\begin{figure}[t]
\includegraphics[width=\linewidth]{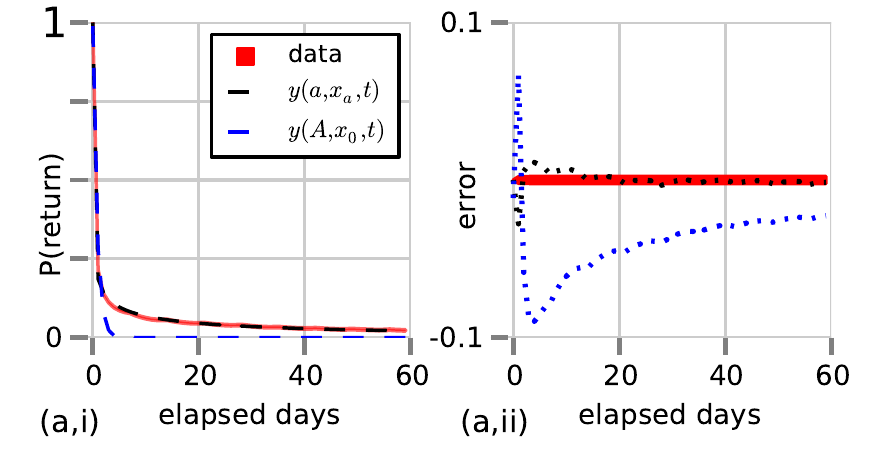}
\includegraphics[width=0.5\linewidth]{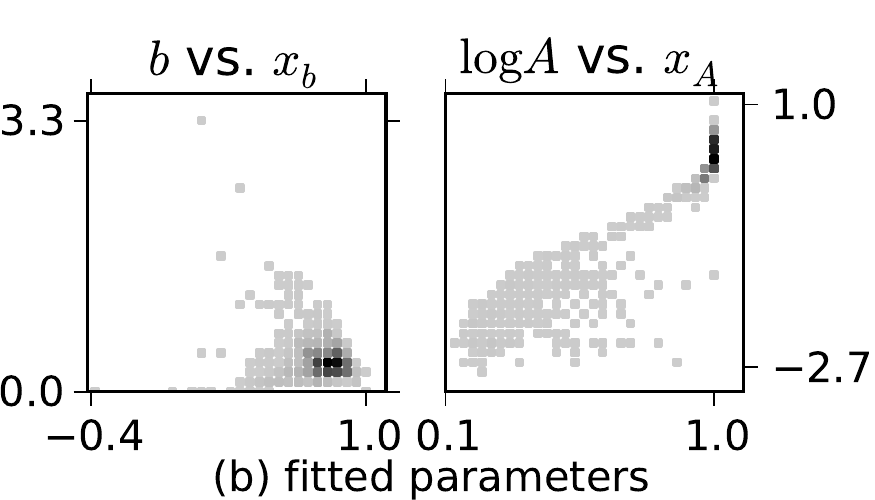}
\includegraphics[width=0.5\linewidth]{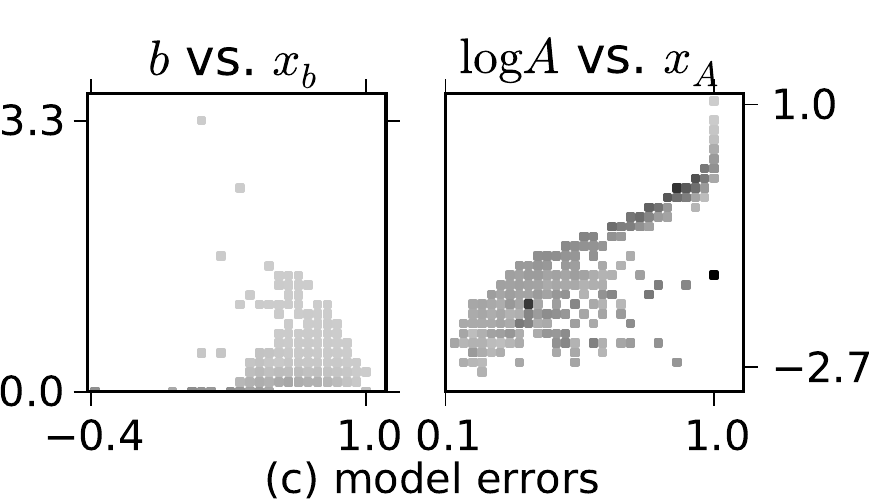}
\caption{Empirical retention data, model predictions, and parameters. \textbf{(a,i):}
Retention data and model predictions for an exemplar app. Error bars on the
data (red solid curve) representing
99.999\% confidence intervals ($4.4172\sqrt{p(1-p)/n}$). \textbf{(a,ii):} Error corresponding
to fit shown in (a,i).
\textbf{(b):} Distribution of fitted retention
model parameters for the apps in \appspop. The shade
represents the frequency of fitted parameter values falling into the given bin
(darker being more frequent). \textbf{(c):} Mean error
achieved by the model for apps with fitted parameters
in the corresponding bin range. The same linear shade scale is used for both
panels, with the lightest gray being $10^{-5}$ and black being $3$ (white
corresponds to no data). $b$ and $x_b$ denote the parameters in the time dependent model,
and $A$ and $x_A$ those in the time indepdendent version.}
\label{fig:retention_curves}
\end{figure}

\begin{align}
\dfrac{dn(t)}{dt} = -x_0 n(t)
\to n(t) = n(0) \exp(- x_0 t),
\label{eqn:exponential_decay}
\end{align}
where $n(t)$ is the number of app users at time $t$.
It turns out that this model does not yield a good fit to the data. However,
the fit improves if we introduce a second parameter to the model by fitting
from the second day; that is, we fit both for the decay rate, $x_0$, and the
fraction of users that returned on day 2. With this relaxation from fitting day
1, the model becomes \begin{align} n(t) = A n(0) \exp(- x_0 t).
\label{eqn:exponential_decay} \end{align}

It is interesting that the exponential decay model fits the day 2 and onward
trend well while not fitting day 1: it is reasonable to expect that there is a
discontinuous transition in the probability of returning given an install
versus an install and a return the following day.
Day 1 users are dominated by those who use the app exactly once, whereas all
the other days contain a signal from users who exhibited at least some level of
continued interest in the app.  Despite its ability to capture this
distinction, this model is unsatisfactory: it entirely ignores the day 1 users,
and even in the two-parameter version it under-predicts retention at long times
(Figure \ref{fig:retention_curves}).

\xhdr{Introducing simple time dependence}

Instead of assuming that people have a constant probability of leaving at every
time step, let us assume that their probability is a simple function of time:
\begin{align}
\dfrac{dn(t)}{dt} = -\dfrac{x_a}{t^a} n(t)
\to n(t) = n(0) \exp\left(- \dfrac{x_a t^{1-a}}{1-a}\right).
\label{eqn:exponential_decay_improved}
\end{align}
This model allows for the possibility that their likelihood of returning
to the app could have a time dependent component, and it
introduces this time dependence with the addition of only one parameter. Notice
that by setting $a=0$ we recover the traditional exponential decay model.

The parameters in this model have an interesting interpretation.
Smaller values
of $a$ indicate that the app users have more momentum, that is, the app has
more sticking power. The parameter $x_a$ is still related to the familiar
probability of depature: small $x_a$ indicates that users are more likely to
continue using the app.

\subsection{Temporal analysis}

After studying the temporal dynamics of
individual apps through our retention model,
we now look for regularities in the temporal patterns
across multiple apps.
We start this
analysis by taking a random sample of all apps, and clustering their time
series of daily active users using a k-means algorithm. All the time series are
normalized by the number of active users on the app's peak day.

\begin{figure}[!b]
\includegraphics[width=0.5\linewidth]{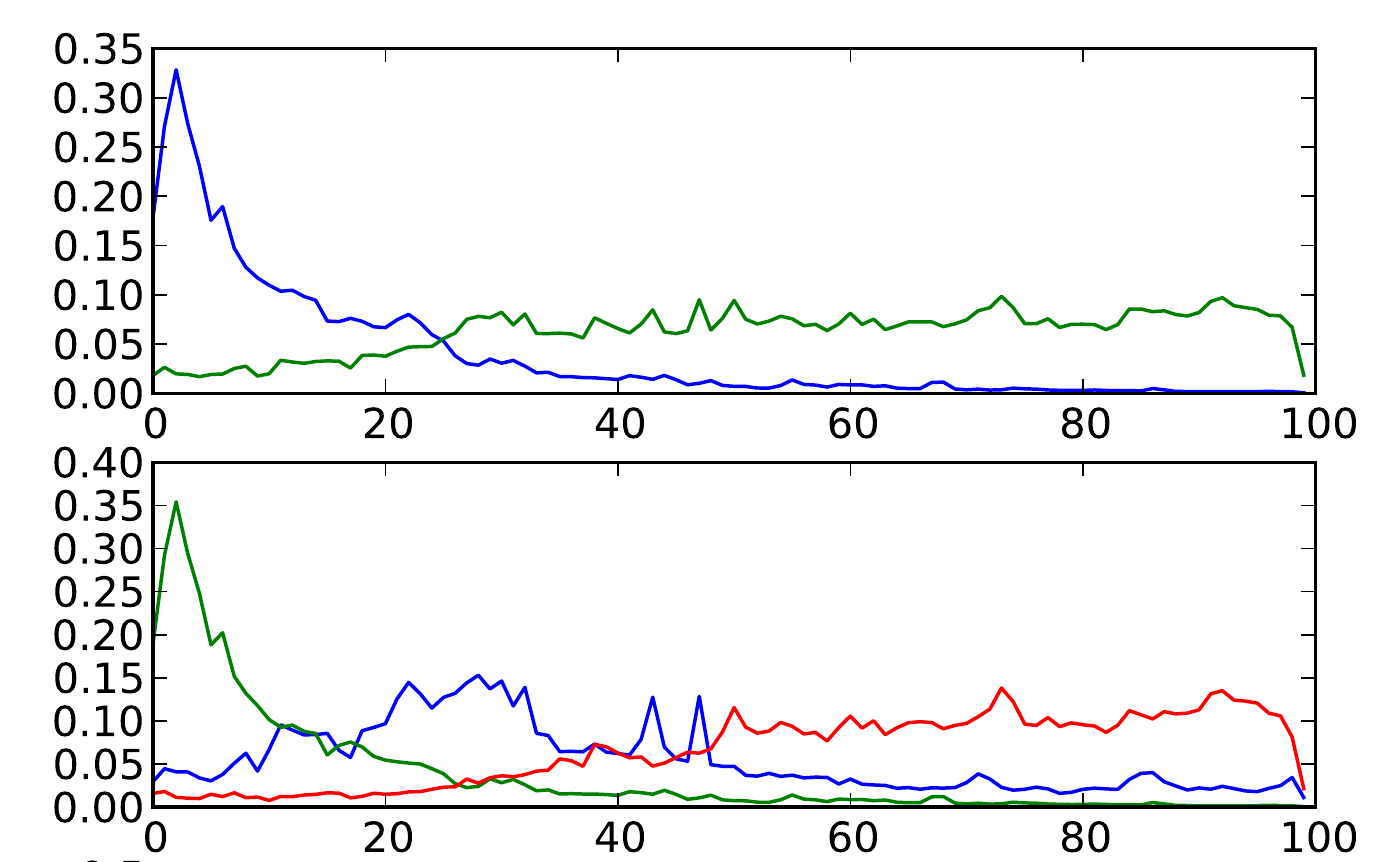}
\includegraphics[width=0.5\linewidth]{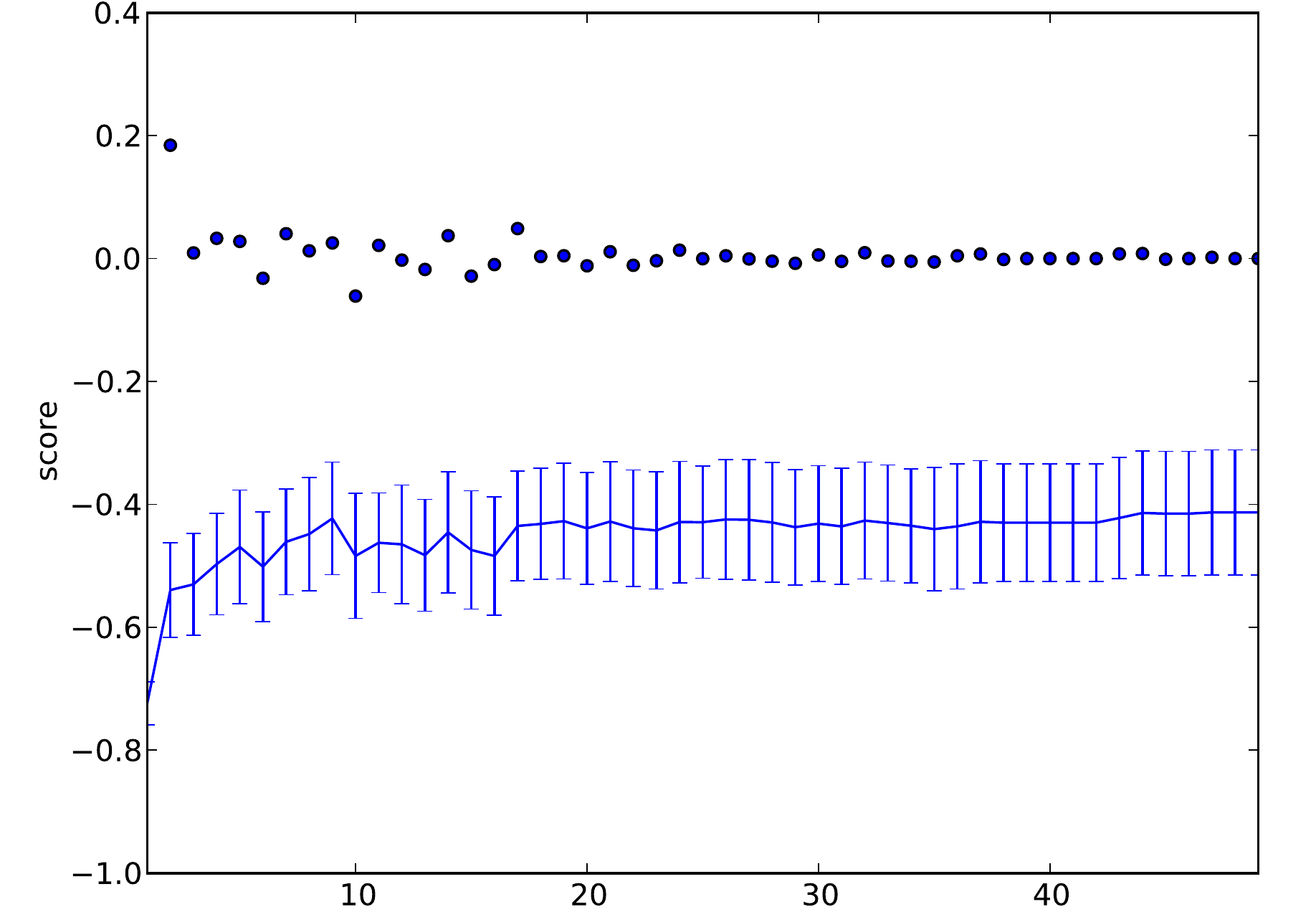}
\caption{\textbf{Left: } $k$-means centroids of DAU for \appsrand, for $k=2$
and $k=3$. The clustering was done on the peak normalized time series of a
100-day observation window. \textbf{ Right:} K-means scores (mean of distances
from nearest centroid) for various values of $k$. Error bars represent 95\%
confidence intervals. Scatter plot is derivative of scores. Notice that no
statistically significant improvement is gained for $k>2$; that is, for all
$k>2$ the score of the resulting clustering is statistically significant
different from that for $k=2$. The scores were evaluated using a 75-25
train-test split, the clusters were generated with 100 restarts, and $L2$ was
used as the distance metric.} 
\label{fig:kmeans}
\end{figure}

By varying $k$, we can get different sets of temporal clusters (see Figure
\ref{fig:kmeans}).
The two clusters generated for $k = 2$ already capture two dominant
temporal trajectories of apps: one
exhibits a clear rise and fall, while the other exhibits a slow but more
sustainable rise. Note that the slow rise shown in the second cluster here may
be misleading: as these apps keep growing and we normalize the time series by
total volume, apps with a bigger user base will appear to have fewer
fluctuations and slow growth/drop. When $k$ increases,  other small temporal
clusters emerge, but they are not significantly different from the two typical
ones.

\begin{figure}
\includegraphics[width=0.5\linewidth]{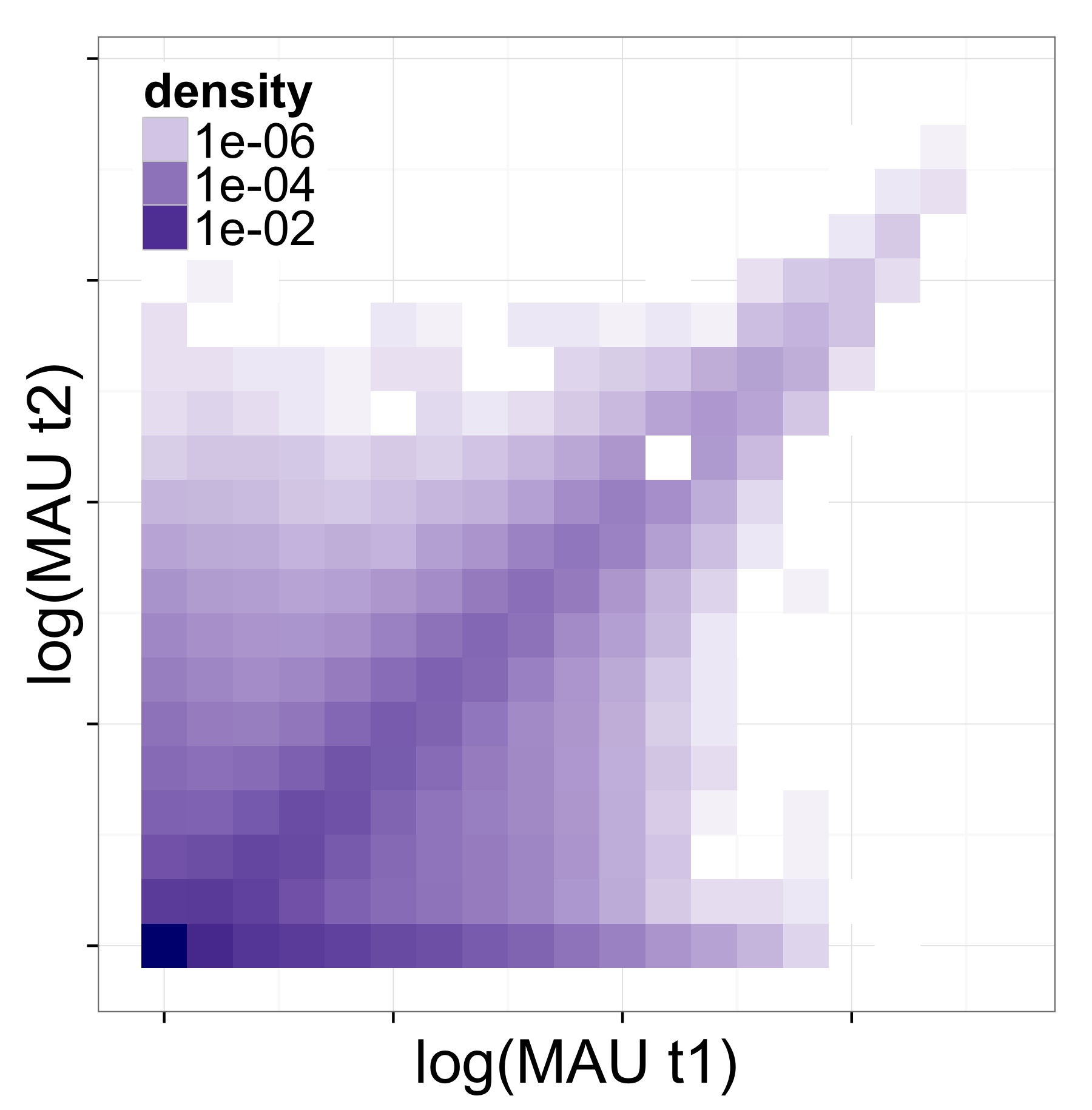}
\includegraphics[width=0.5\linewidth]{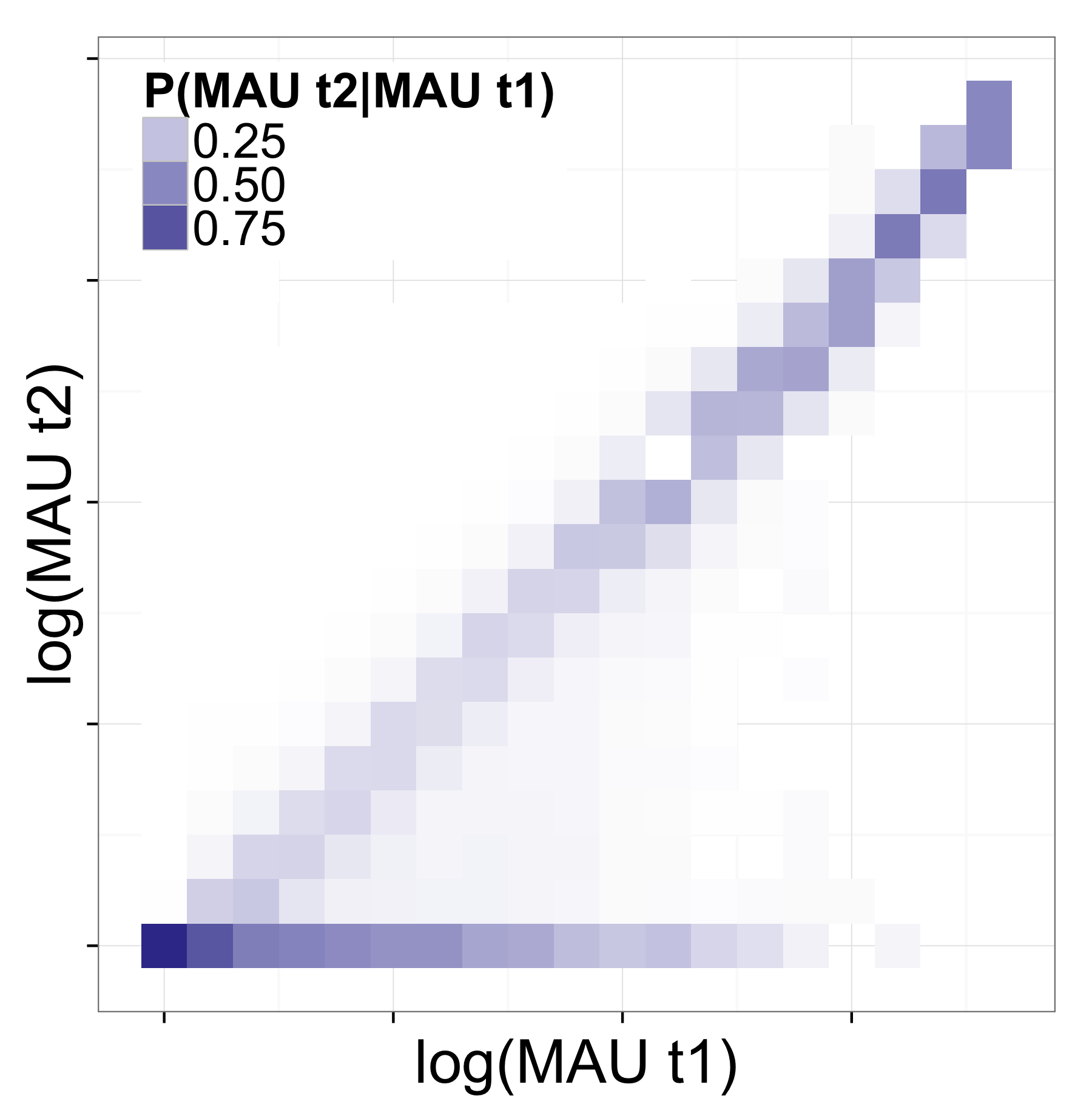}
\caption{MAU of all apps in June 2013 (horizontal) and June 2014 (vertical).
\textbf{ Left:} Number of apps with MAU@$t_1$ and MAU@$t_2$ corresponding to the
specified bins. \textbf{ Right:} P(MAU@$t_2$| MAU@$t_1$) is the empirical
probability of an app having $y$ users at $t_2$ given that it had $x$ users at $t_1$.
At $t_2$ we observe that apps tend to either continue at the same level of popularity
as they experienced at $t_1$ (bright diagonal) or exhibit a dramatic decrease in popularity
(bright band on horizontal axis). Apps that are more populare have greater rates of continued success.
However, when their popularity drops, the collapse tends to be complete.
} \label{fig:twopointmau} \end{figure}

Then, for all apps that existed on June 1 2013, we compute their monthly
active users (MAU) on that day and one year out. We plot those distributions
against each other in Figure \ref{fig:twopointmau}, where the color
indicates the number of apps with the stated pairing of MAUs. The
right-hand
figure normalizes the columns, so that each bin column can be interpreted as
the probability of ending up at the indicated MAU,
given the current position.  We can
see that apps are most likely to stay at approximately the same MAU, but that,
especially for very popular apps, there is a subpopulation that loses almost
all of their users.  This pattern suggests a natural underlying binary
prediction problem: given that an app enjoys current success will it continue
to be as popular one year later?


\section{Predicting app success}
\label{sec:prediction_setup}

In the previous sections we have seen that apps can be described in a variety
of ways. We began by exploring the relationship between an individual's social
network and their likelihood of adopting an app, and in general how app usage
is clustered in the social network. We also related a user's likelihood of
adopting an app to their individual characteristics relative to those of the
current app's users.  Next we observed that, though overall patterns of
adoption can be quite complex, an app's retention properties are well described
by a simple model with a small set of
parameters. And finally we again saw that while
the fine grained activity level for any app is complicated, in the long term
apps tend to either continue at the same activity level or diminish in
popularity.

In each analysis we considered either hundreds or thousands of popular apps
from this ecosystem, and saw that these various low dimensional features had
interesting and diverse distributions across the population.

This brings us to our final set of questions:
can we use an app's social, demographic,
retention, and temporal features to predict whether or not it will be successful
in the long term?


\begin{table*}[htbp!]
\begin{tabular}{|l|m{5in}|}
\hline
\hline
\multicolumn{2}{|c|}{ Temporal} \\
\hline
med /$\min$/$\max$ DAU$_{mo.X}$ & median, min, max number of daily users in month $X$ of observation\\
med /$\min$/$\max$ $\Delta$DAU$_{mo.X}$ & median, min, max of change in daily users within month $X$ (DAU$_{X} - $DAU$_{X-1}$) \\
med /$\min$/$\max$ $\Delta^2$DAU$_{mo.X}$ & median, min, max of second order change in daily users within month $X$  \\
med /$\min$/$\max$ DAU$_{year}$ & median, min, max of $DAU_{X}$ for $X\in 1, \dots, 12$ \\
med /$\min$/$\max$ $\Delta$ DAU$_{year}$ & median, min, max of $\Delta DAU_{X}$ for $X\in 1, \dots, 12$ \\
med /$\min$/$\max$ $\Delta^2$ DAU$_{year}$ & median, min, max of $\Delta^2 DAU_{X}$ for $X\in 1, \dots, 12$ \\
$\Delta_{year}$ DAU & med DAU$_{12}$ - med DAU$_{1}$ \\
$\Delta_{year}\Delta$ DAU & med $\Delta$ DAU$_{12}$ - med $\Delta ^2$ DAU$_{1}$ \\
$\Delta_{year}\Delta^2$ DAU & med $\Delta^2$ DAU$_{12}$ - med $\Delta^2$ DAU$_{1}$ \\
*WAU, MAU, users, new users & Same statistics as listed for DAU above, considering instead weekly users, monthly users, total users, and new users\\
\hline
\hline
\multicolumn{2}{|c|}{ Demographic} \\
\hline
Country$_{X}$ / $P($Country$_{X})$& Number, fraction of users from country $X$ \\
Gender$_{X}$ / $P($Gender$_{X})$& Number, fraction of users who stated their gender to be $X$ \\
Age$_{X}$ / $P($Age$_{X})$& Number, fraction of users who stated their age to be $X$ \\
$l_{k,7}$ / $P(l_{k,7})$& Number, fraction of users who were active on Facebook for $k$ out of 7 days\\
is30 / isnot30 / $P(is30)$& Number of users who are / aren't monthly active Facebook users; fraction of users who were monthly active Facebook users\\
Entropy(Country) & Entropy of country user distribution: $-\sum_{X\in \text{Countries}} P(\text{Country}_X) \log_2 P(\text{Country}_X)$\\
Entropy(Gender) & Entropy of gender user distribution: $-\sum_{X\in \text{Genders}} P(\text{Gender}_X) \log_2 P(\text{Gender}_X)$\\
Entropy(Age) & Entropy of age user probability distribution: $-\sum_{X\in \text{Ages}} P(\text{Age}_X) \log_2 P(\text{Age}_X)$\\
Entropy($l_7$) & Entropy of $l_7$ distribution: $-\sum_{k\in 1,\dots,7} P(l_{k,7}) \log_2 P(l_{k,7})$\\
Entropy(is30) & Entropy of is30 distribution: $-[P(is30) \log_2 P(is30) + (1 - P(is30)) \log_2 (1-  P(is30))] $ \\
\hline
\multicolumn{2}{|c|}{ Retention} \\
\hline
N(t) &  Number of users who returned $t$ days after their first login  \\
P(t) &  Empirical probability of a user returning $t$ days after their first login  \\
$a$, $x_a$  & Parameters for best fits of time dependent model: $N(t) = N(0) exp\left[\dfrac{- x_a t^{1-a}}{ {1-a}}\right]$\\
$A$, $x_0$  & Parameters Least squares parameter fits of time independent model:$N(t) = A N(0) exp[- x_0 t]$\\
\hline
\multicolumn{2}{|c|}{ Social} \\
\hline
med / $\max$ deg & median and maximum number of friends of an app user \\
med / $\max$ using & median and maximum number of friends of an app user who also use the app \\
$p(x | y)$ & \tworows{sociality: empirical probability of having adopted an app given that a friend has, }{i.e. mean fraction of an app user's friends who also use the app} \\
$p(x | y)/ p(x) $ & relative change in probability of a user adopting an app given that their friend has \\
\hline
\multicolumn{2}{|c|}{ SIRS model} \\
\hline
$S_0$ & susceptible population size, i.e. number of Facebook users who are interested at this app\\
$\alpha$  & probability of a non-user adopting the app due to non-social reasons\\
$\beta$ & probability of a non-user adopting the app through social process\\
$\gamma$ & probability of active user becoming inactive\\
$\epsilon$ & probability of in-active user being drawn back by active users\\
$pred(day_k)$ & the DAU prediction at day $k$ for $k$ between 2013-06-01 and 2014-06-01 \\
\hline
\hline
\end{tabular}
\caption{Features used for training and testing the binary app success
prediction tasks. Features were measured for all apps in \appspop\ (see Table
\ref{tab:data}), with the exception of the SIRS model features due to issues of
model convergence.}
\label{tab:features}

\end{table*}

\subsection{Predicting the longevity of apps}
Note that we have seen empirically that the question of
an app's long term success is well approximated by a binary variable (see
Figures \ref{fig:kmeans} and \ref{fig:twopointmau}).
In this subsection and the next, we will consider two variations on a binary prediction task.
One task is straightforward: given a collection of promising apps, we want
to  predict which apps will have persistent success over the next year. The
other task is based on a pairwise evaluation:
to compare a pair of similarly popular apps and predict which one
will be more successful in the future.

First, we consider the task of predicting which apps in the entire population
will continue to be successful. Based on the number of active users on June
2014, we label an app as a positive example if it has over 50\% of the number
of active users it had in June 2013, and we label it as a negative example if
has lost more than 50\% of its users.
This labeling turns out to provide us with a balanced class
distribution, with the guess-all-positive baseline being 50\%.  For this binary
classification task we built and evaluated the model by training random forests
on apps in \appspop, where each app is represented as a vector of the features
in Table \ref{tab:features}.

The prediction performance results are shown in Table
\ref{tab:prediction}, and the use of all the available features
leads to performance above 70\% on this binary task.
We find that the temporal features are the best single set of features, with
the most important features being the median number of users in months 8 and 9
of the 12 month observation period (June 1 2012-June 1 2013 -- see Table
\ref{tab:data}). The apps that would continue to be successful also had a higher weekly minimum; given that the overall popularity of the apps between classes is evenly distributed,
we interpret this high weekly minimum as a signal of stability, and that this stability was a positive predictor.

Individual user attributes yielded the second highest
performance, with the most important class features being activity-based
ones: $l_{5,7}$ and $l_{6,7}$ (
$l_{k,7}$ is the fraction of app users that were also active Facebook users for
past $k$ out of 7 days). We observe that for $k=0,\dots,6$, negative examples
are correlated with greater values of $l_{k,7}$, whereas for $l_{7,7}$, the trend reverses,
and the positive examples with more users who are active on Facebook every day. This means that having users
who were also highly active Facebook users is a positive indicator of success.

Among all the retention features, the most important one was the fitted parameter $x_A$, which
represents the ``departure probability'' in the exponential decay model of users
leaving an app. Not surprisingly, we find that the positive examples tend to
have lower $x_A$ than negative examples, indicating that having users who
continue using the app for an extended period of time (i.e. a lower leaving
probability) is correlated with the app's long-term success.

Finally, the most important structural features were sociality, i.e., average
user degree, and mean/max number of friends who used the app. For the latter
two we could not notice any significant differences between the two classes,
but we do notice that high sociality is a negative indicator of success. This
is likely due to the fact that we normalize the sociality measure ($p(x|y)$) by
the popularity of the app ($p(x)$); thus those apps with very high sociality
score are relatively small, and tend to be the ones situated in a very
specific, niche market. Indeed, we find that if we consider the separate
distributions of the numerator and denominator, we observe that $p(x|y)$ is
indistinguishable for the two classes, while $p(x)$ is a positive indicator,
leaving $p(x|y)/p(x)$ as a negative indicator.

\xhdr{SIRS model}

In general the task of
predicting an app's time-series trajectory is a rich and interesting problem,
but the binary nature of trajectories that we observed motivated our
simplification to the binary prediction task.
To explore the potential inherent in modeling richer properties of
the time series,
we also consider a model of app usage via
a set of interacting reaction diffusion
processes, much like a chemical reaction. The model we use was proposed by
Ribeiro~\etal~\cite{Ribeiro:2014}, and falls into the well-known class of SIRS
models. We will briefly describe how we implemented this model, and when we
return to our underlying prediction task, we will consider the predictions
and parameters from this model as an additional set of features.

Ribeiro~\etal~\cite{Ribeiro:2014} proposed a model
describing the dynamics of a webservice of daily activity time series, derived
from the classical epidemic model and comprised of a set of reaction diffusion
processes. The model is specified by a set of parameters, including the
estimate of the susceptible population, and the transition probabilities
between different states. Ribeiro also outlines a framework for fitting these
parameters given a window of time series activity levels, and then uses them to
extrapolate and make a long term prediction of future activity levels.
We implemented a model very similar to the one described
in~\cite{Ribeiro:2014}. We fitted the model using a Monte Carlo process using
time series from June 2, 2012 to May 25, 2013 (the same period from which we
extract temporal features), and used the fitted model to generate predictions
between May 26, 2013 and May 15, 2014.

There are two things we note about the SIRS model. First, as we try to
predict the future of apps from a fixed time point, the apps we are studying
can be in very different life stages. For example, some apps in our dataset had
only existed for a short period of time by the observation day, and thus have very
limited time series data to compute a good fit of the SIRS model. Second, some
underlying assumptions in the SIRS model, such as the constant rate of user
adoption through advertisement or word-of-mouth process, may not hold in reality. As a
result, the model would not converge for certain apps, especially the ones that
experienced large fluctuations in their lifecycles.

Nevertheless, we were able to fit over two-thirds of the apps in \appspop.
Among them, 1100 apps had reasonable convergence and error estimates. We then
used both the fitted parameters and the predicted time series as our features
for this subset of 1100 apps.  On that subset of 1100 apps, the relative
performance of the other features sets was the same (all combined features
yield the highest performance, followed by temporal, then demographic,
retention, and finally social).

We find that the features from the SIRS model perform
worse than the retention features but better than the social features.
Thus, despite the richness of the time-series modeling made possible
by the SIR framework, as a feature set it does not perform as well
as other measures incorporating temporal properties, including the
retention model from the previous section.


\begin{center}
\begin{table}[htpb]
\begin{small}
\begin{tabular}{|C{0.62in}|C{0.40in}|C{0.24in}|C{0.26in}|p{1.1in}|}
\hline
\hline
\emph{Feature set} &\hspace{-4pt} \emph{accuracy} & \emph{\tworows{prec:}{+;-}} & \emph{\tworows{recall:}{ +;-}} & \emph{\tworows{\tworows{top 2 features:}{\{among all\};}}{\{within class\} }}\\
\hline
Baseline &0.50 &&&  \\
\hline
All& 0.73& \tworows{0.72;}{0.74 } & \tworows{0.74;}{0.72}& \tworows{\tworows{med users$_8$}{med users$_9$};}{--}  \\
\hline
Temporal&  0.71& \tworows{0.72;}{0.7 } & \tworows{0.68;}{0.74}& \tworows{\tworows{$\Delta_{year} WAU$,}{$\min WAU_{11}$};}{\tworows{med users$_8$}{med users$_9$}}\\
\hline
Demographic& 0.66& \tworows{0.64;}{0.68 } & \tworows{0.70;}{0.61}& \tworows{$l_{6,7}$, $l_{5,7}$;}{$l_{6,7}$, $l_{5,7}$}\\
\hline
Retention& 0.61& \tworows{0.59;}{0.64 } & \tworows{0.70;}{0.53}&\tworows{ $x_a$, $x_A$;}{day 2 and 3 returns}\\
\hline
Social & 0.6& \tworows{0.59;}{0.61 } & \tworows{0.60;}{0.59}&\tworows{$\dfrac{p(x|y)}{p(x)}$, $\langle$user degree$\rangle$;}{\tworows{Mean and max \# of}{ friends using the app}} \\
\hline
\hline
\end{tabular}
\end{small}
\caption{Prediction performance results for five combinations of features.
Precision and recall: top and bottom rows are for positive and negative
classes, respectively. Features are ranked by out-of-bag importance estimates
while
training the random forests.
We trained the classifier using all the features, and report the
most important ones in each category in the top row (``among all''), and train
the classifier with only the features in each category, and report the top
opens in lower row (``within class'').}

\label{tab:prediction} \end{table} \end{center}

\subsection{Predicting pairwise relative success}
\label{sec:pairwise}

Next we formulate a separate but related prediction task, by constructing a
pairwise comparison version of predicting app success.  Given that two apps
have approximately the same monthly active users at $t_1$ (MAU@$t_1$), and by
$t_2$ they had diverged from each other, we want to predict at time $t_1$ which
app is going to be more successful.  We evaluate this problem with a variety of
thresholds for what we considered ``near-'' and ``long-''term predictions of
MAU. This prediction task is particularly useful when investigating a set of
competitive apps in the same market.  Intuitively, it is difficult to tell
similar apps apart at an early stage~\cite{ribeiro2014modeling}. However, by
looking at pairs whose outcomes at $t_2$ are successively farther apart, we can
control for the difficulty of the task and understand when it becomes feasible
to predict such divergence.


For the pairwise prediction task we begin by generating a 50-50
train-test split
between apps, and represent each pair of apps as a concatenation
of two feature vectors, again using the features from Table \ref{tab:features}.
We then introduce a subtle variation
to make the setup more relevant to a real-world
scenario. The features and labels used in the training stage are generated
using snapshots of our datasets at $t_0$ and $t_1 = t_0 + 6$ months, while those
used for testing are generated using snapshots at $t_1$ and $t_2 = t_1 + 6$
months.  This simulates the practical scenario of observing the app population
at $t_1$, learning which characteristics of apps lead to their success, and using the learned knowledge to predict the future.

\begin{figure}
\includegraphics[width=\linewidth]{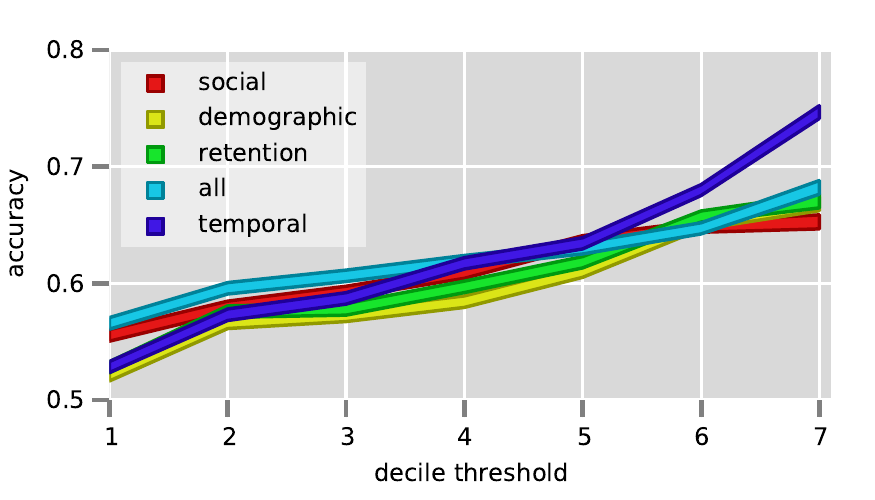}
\caption{Prediction accuracy for the pairwise relative success prediction task, as a function of decile threshold, $k$.
}
\label{fig:pairwise}

\end{figure}

Two apps are considered to start off as being ``comparable'' if they fall into
the same decile at $t_{0/1}$ (train/test), and are considered ``distinct''
if they
are at least $k$ deciles apart at $t_{1/2}$ (train/test). In Figure \ref{fig:pairwise}
we see that prediction accuracy increases monotonically with $k$, and that the
best set of features (temporal) ultimately yield 75\% prediction accuracy. The
other most striking feature of Figure \ref{fig:pairwise} is that for most of
the threshold window, all the features yield approximately the same
performance. Each set of features, besides demographic, takes a turn at being
both the top performer and the lowest. The individual feature analysis that we
did was consistent with the observation that this task is not highly sensitive
to the choice of features. To analyze which features could best discriminate between
positive and negative examples we used the two-sided
Kolmogorov-Smirnov test to compare the distributions of each feature for
positive and negative examples. We find that, with the exception of a few
underpopulated demographic features, the Kolmogorov-Smirnov test finds that
each feature is distinguishable between the negative and positive
examples with $p$-values extremely close to zero.
\reminder{Consider adding figure showing distribution for two interesting features,
choose somewhat arbitrarily since they are almost all significant.}
\reminder{Concluding remarks - Maybe mention how it is interesting this task
does so well given that the features from one time period are being asked to
generalize to predict in another time period, and that they continue to do
well.}

\section{Related Work}

Sociologists and economists have long studied the problem of product adoption
and retention.  Early work in this domain focused on the diffusion of
innovations, as people proposed a series of mathematical models to describes
the adoption of new products by consumers, such as the ``S-shaped'' adoption
curve~\cite{Tarde:1903} and the Bass model~\cite{Bass:1969}. These models have
been successful in predicting the impact of advertising, especially the effect
of advertising through mass-media and billboards.  Other work has focused on
the diffusion of innovations and products through social
ties~\cite{Rogers:1962}.  With the rise of social media and online social
networks, there has been more and more evidence that the social influence, i.e.
the word-of-mouth effect, is playing a increasingly important role at driving
the adoption of products and services
~\cite{Leskovec:2006:ec,muchnik2013social,bakshy2012role}.

To understand how products and information spread in social networks, most existing work tries to predict
the volume of popularity, such as the the size of online communities~\cite{Backstrom2006group},
the number of fans of Facebook pages~\cite{Sun:2009}, and the usage of hashtags on Twitter~\cite{Bakshy:2011,Romero:2011,Weng:2013}.
While these work showed the correlation between the scale of diffusion and its structural and topical properties, there has been
a recent line of work questioning the predictability of large viral events~\cite{Salganik:2006,Bakshy:2011}.
In response, Cheng \etal~\cite{Cheng:2014} showed that it is possible to predict how much more a cascade
will grow by observing the temporal and other features of its spread up to the present time.

Besides being a key predictor for cascade size, the temporal dynamics of cascades have been an
interesting research topic~\cite{Leskovec:2009,Crane:2008,Yang:2011}. Upon the discovery of several
robust temporal classes of cascades on different platforms, most studies on the temporal dynamics
focused on bursty events~\cite{Barabasi2006}, or the peak volume~\cite{Crane:2008,Leskovec:2009}.
Indeed, the majority of popular things spread on-line enjoy very short attention span: the
popularity rises and drop quickly, usually within a few hours or a day~\cite{Yang:2011,Wu:2011}.
The persistence of interest, although rare, is rather intriguing. Wu \etal~\cite{Wu:2011} found that the longevity
of URLs on Twitter can be explained by the intrinsic cultural value of the content they link to. Follow-up work
showed that information with positive sentiment is more likely to persist~\cite{Wu:2011longevity}.
Ducheneaut \etal. discovered that smaller and denser guilds in World
of Warcraft are more likely to survive longer~\cite{Ducheneaut:2007}.

While many papers correlate the temporal patterns of cascades with its empirical properties, some researchers
have developed theoretical models on individuals' choice of adopting and engaging with a product or
activity~\cite{Barabasi2006,Ribeiro:2014}. These models are useful at depicting the mechanism behind
the observed temporal dynamics, however, it is unclear how generalizable they are beyond the particular product or activity studied.

Our work contributes to current research in two major aspects.

First, we study the entire lifecycle of apps over a timespan as long as 5 years \reminder{Is this true?
I thought we focused primarily on a 2 year period?}. Our focus is the persistency of growth other than
the peak popularity. Different from a viral YouTube video or a meme photo, successful apps needs to
engage with their users repeatedly. Therefore, we spent a significant amount of work analyzing and
modeling the retention of apps, and showed its importance to the long-term success of apps.

Second, we study thousands of apps at once. Previously, most papers examined the adoption and
retention of a single product/activity, thus their results might not be generalizable to other domains.
By studying a large selection of apps on Facebook, we are able to control for app-specific features
and understand how the characteristics of an app interact with its social and temporal dynamics.

Some work similar to ours includes a recent study of the growth and longevity
of online communities~\cite{Kairam:2012}, the modeling and prediction of the
temporal pattern of membership-based websites~\cite{Ribeiro:2014}, and a study
of mobile app adoption over a small real-life social network~\cite{Pan:2011}.
Our study builds on these papers in both the scope and the variety of examples
examined. Also, with the rich dataset we have about apps, users, and the
underlying social graph, we are able to introduce several new theoretical
and analytical models, and to compare them with recent formalisms
\cite{Ribeiro:2014}.
By incorporating the
parameters of a fitted model as part of the feature set, we are able to extend
and compare different methodologies.

\section{Conclusion}
In this paper we studied the lifecycle of apps: as they grow and thrive, and, in some cases, as they decline.
We studied differences in their development, looking for clues to their future fate. First, we
sought parameters with which to model the interaction between the app and the individual.
We found that a simple exponential decay, even with an adjustment for attrition after the first
day of use, did not accurately capture user retention. Instead, those who keep using the app
over a longer time period are less and less likely to stop. Modeling retention of individuals in
this way is helpful in predicting app success.

Another dimension goes beyond the individual to whether the app is
adopted socially.  Apps vary widely in the sociality of their
adoption, and we find heterogeneity in the apps based on
how their adoption probabilities depend on the connectedness of
friends who use the app and the similarities in attributes
between an adopter and his or her friends.

The features most predictive of an app's future dynamics are those
describing its past growth trajectory. More widely adopted apps that
have recently been on a growth trajectory are more likely to persist.
Given a range of features, we obtain over
20\% absolute improvement over random
guessing when it comes to making a binary prediction as to sustained
activity for an app.
We also obtain strong performance
when we formulate the problem as one of
matching two apps of roughly equal size which take different
trajectories, and trying to distinguish the two with a much higher
than random accuracy.

There are a number of further aspects of the app ecosystem that
would be interesting to take into account in future work.
First, app adoption is driven in part by
the marketing and other recruitment strategies of the app owners.
Although our models incorporate the numbers of new users coming to the
app over time, they do not differentiate between organic growth and
advertising-driven growth.  Furthermore, it is not clear whether
sociality of apps might accelerate growth or decline or both. Finally,
it is unclear whether some features might be early harbingers of
future behavior, e.g. whether the change in retention of long-time or
recently acquired users is more useful in forecasting the eventual
adoption of the app. We leave these and other questions for future
work.
\vspace{15pt}

\bibliographystyle{plain}
\bibliography{refs}

\end{document}